\documentclass[prl,twocolumn,superscriptaddress]{revtex4}
\usepackage{xcolor}
\usepackage{graphicx}
\usepackage{amsmath,amssymb}

\usepackage[normalem]{ulem} 
\usepackage{color} 

\begin{document}

\title{Testing of Generalized Uncertainty Principle With Macroscopic Mechanical Oscillators and Pendulums}

\author{P. A. Bushev}
\affiliation{Experimentalphysik, Universit\"{a}t des Saarlandes, D-66123 Saarbr\"{u}cken, Germany}

\author{J. Bourhill}
\affiliation{ARC Centre of Excellence for Engineered Quantum Systems, University of Western Australia, Crawley,
Western Australia 6009, Australia}

\author{M. Goryachev}
\affiliation{ARC Centre of Excellence for Engineered Quantum Systems, University of Western Australia, Crawley,
Western Australia 6009, Australia}

\author{N. Kukharchyk}
\affiliation{Experimentalphysik, Universit\"{a}t des Saarlandes, D-66123 Saarbr\"{u}cken, Germany}

\author{E. Ivanov}
\affiliation{ARC Centre of Excellence for Engineered Quantum Systems, University of Western Australia, Crawley,
Western Australia 6009, Australia}

\author{S. Galliou}
\affiliation{FEMTO-ST Institute, Universit\'{e} of Bourgogne Franche-Comt\'{e}, CNRS, ENSMM, 25000 Besan\c{c}on, France}

\author{M. E. Tobar}
\affiliation{ARC Centre of Excellence for Engineered Quantum Systems, University of Western Australia, Crawley,
Western Australia 6009, Australia}

\author{S. Danilishin}
\affiliation{Institut f\"{u}r Theoretische Physik, Leibniz Universit\"{a}t Hannover and Max-Planck Institut
f\"{u}r Gravitationsphysik (Albert-Einstein-Institut), 30167 Hannover, Germany}

\date{\today}

\begin{abstract}
Recent progress in observing and manipulating mechanical oscillators at quantum regime provides new opportunities of studying fundamental physics, for example to search for “low energy signatures of quantum gravity”. For example, it was recently proposed that such devices can be used to test quantum gravity effects, by detecting the change in the $[\hat{x},\hat{p}]$ commutation relation that could result from quantum gravity corrections. We show that such a correction results in a dependence of a resonant frequency of a mechanical oscillator on its amplitude, which is known as amplitude-frequency effect. By implementing of this new method we measure amplitude-frequency effect for 0.3 kg ultra high-Q sapphire split-bar mechanical resonator and for $\sim10^{-5}$~kg quartz bulk acoustic wave resonator. Our experiments with sapphire resonator have established the upper limit on quantum gravity correction constant of $\beta_0$ to not exceed $5.2 \times 10^6$, which is factor of 6 better than previously measured. The reasonable estimates of $\beta_0$ from experiments with quartz resonators yields $\beta_0<4\times10^4$. The data sets of 1936 measurement of physical pendulum period by Atkinson \cite{Atkinson1936} could potentially lead to significantly stronger limitations on $\beta_0 \ll 1$. Yet, due to the lack of proper pendulum frequency stability measurement in these experiments the exact upper bound on $\beta_0$ can not be reliably established. Moreover, pendulum based systems only allow to test a specific form of the modified commutator that depends on the mean value of momentum. The electro-mechanical oscillators to the contrary enable testing of any form of generalized uncertainty principle directly due to a much higher stability and higher degree of control. 

\end{abstract}



\maketitle
\Roman{section}

\section{I. INTRODUCTION}
 At present, one of the grandest challenges of physics is to unite its two most successful theories: quantum mechanics (QM) and general relativity (GR), into a single unified mathematical framework. Attempting this unification has challenged theorists and mathematicians for several decades and numerous works have highlighted the seeming incompatibility between QM and GR~\cite{HossenfelderBook}. It was generally supposed that this required energies at the Planck scale and therefore beyond the reach of current laboratory technology~\cite{AmelinoCamelia2001}. However in the relatively recent publication, I. Pikovsky et al.~\cite{Pikovski2012} proposed a new method of testing a set of quantum gravity (QG) theories~\cite{Maggiore1993,Maggiore1994,Kempf1995,Garray1995,Vagenas2009} by using ingenuitive interferometric measurement of an optomechanical system. The prediction of most of the QG theories (such as, \textit{e.g.}, string theory) and the physics of black holes lead to the existence of the minimum measurable length set by the Planck length $L_p=\sqrt{\hbar G/c^3} \simeq 1.6 \times 10^{-35}$~m~\cite{Maggiore1993,Garray1995,Vagenas2009}. This results in the modification of the \textit{Heisenberg uncertainty principle} (HUP) in such a way as to prohibit the coordinate uncertainty, $\Delta x\sim\hbar/\Delta p$, from tending to zero as $\Delta p\to \infty$ ~\cite{Hawking1978,Scardigli1999,Scardigli2010,AmelinoCamelia2000,vanDam2000}. The modified uncertainty relation, known as \textit{generalised uncertainty principle} (GUP), is model-independent and can be written for a single degree of freedom of a quantum system as:
\begin{equation}\label{eq:GUP}
\Delta x\Delta p \geq \frac{\hbar}{2}\bigg[1+\beta_0\frac{\Delta p^2+\langle p\rangle^2}{M^2_p c^2}\bigg]\,,
\end{equation}
where $\beta_{0}$ is a dimensionless model parameter, $M_p =\sqrt{\hbar c /G} \simeq 2.2 \times 10^{-8}$~kg is Planck mass and $\langle p\rangle$ is the quantum ensemble average of the momentum of the system. The dependence of minimum uncertainty of coordinate on average momentum is questionable but some theories~\cite{Kempf1995,Vagenas2009} explain as reflects the connection of spacetime curvature and the density of energy and matter manifested in Einstein's equations of general relativity.

Other more intuitive form of the GUP, \textit{e.g.},
\begin{equation}
\Delta x \Delta p > \frac{\hbar}{2}\bigg[ 1+\gamma_0 \bigg(\frac{\Delta p}{M_p c} \bigg)^2\bigg],
\label{GUP}
\end{equation}
which depends only on the uncertainties of the canonical variables of the particle but not on their mean values is predicted in~\cite{Garray1995}. To test this theory one need to either measure the deviation of the oscillators ground-state energy $E_\text{min}$ with respect to its unperturbed value $\hbar\Omega_0/2$~\cite{Marin2012}, or to test of QG corrections to the dynamics of the quantum uncertainty of the mechanical degree of freedom using pulsed measurement procedure proposed in~\cite{Pikovski2012}, which requires quantum level of precision. 


The lowest modal energies measured in large mechanical systems such as AURIGA detector with effective mass of the mode $m_{\text{eff}} \simeq 1000$~kg~\cite{Marin2012} and in dumbbell sapphire oscillator with $m_{\text{eff}}\simeq 0.3$~kg~\cite{Bourhill2015} set the limit on the QG model parameter $\gamma_0 \lesssim 3\times 10^{33}$, which is still too large compared to the predicted values of the order of unity~\cite{Das2008}.

\section{II. THEORY}

From the GUP \eqref{eq:GUP} one can derive the new canonical commutation relation:
\begin{equation}
[\hat{x},\hat{p}]_{\beta_0}=i\hbar \bigg[1+\beta_0 \bigg(\frac{\hat p}{M_p c}\bigg)^2\bigg],
\label{commutator1}
\end{equation}
that is deformed by the QG correction defined by the model parameter $\beta_0$. As shown by Kempf et al.~\cite{Kempf1995}, parameter $\beta_0$ defines the scale of the absolutely smallest coordinate uncertainty $\Delta x_{min} = \hbar\sqrt{\beta_0}/(M_{p}c)$. In this work, we experimentally set an upper limit on the value of the model parameter $\beta_0$ using the dynamical implications of the contorted commutator on the oscillations of a high-Q mechanical resonator of mass $m$ and (unperturbed) resonance frequency $\Omega_0$.

We start our consideration with the simple premise that the modification of the fundamental commutator for a harmonic oscillator is equivalent to the nonlinear modification of the Hamiltonian by means of the perturbative transformation of momentum, $\hat p \rightarrow \hat p-\beta_0 \hat p^3/(3M_{p}^2 c^2) $, which restores the canonic commutator, $[\hat{x},\hat{p}]=i\hbar$, at the expense of adding the non-linear term to the Hamiltonian of the resonator: $\hat H \rightarrow \hat H_0 + \Delta \hat H = \bigg( \hat p^2/2m +m\Omega_0^2\hat x^2/2 \bigg)+\beta_0 \hat p^4/\big(3m{(M_p c)}^2\big)$. Such non-linear correction results in the dependence of the oscillator resonance frequency on its energy~\cite{Kempf1995,Vagenas2009,Bosso2017}. The dynamics of the system can be described by a well known Duffing oscillator model characterized by amplitude dependence of the resonance frequency, i.e. so called amplitude-frequency effect~\cite{Gufflet2001,Marin2015}. The necessary frequency resolution in order to sense subtle QG effects can be estimated by using the following expression: 
\begin{equation}
\delta \Omega(A) /\Omega_0 = \beta_0 \big( m_{\text{eff}} \Omega_0 A / M_p c \big )^2,
\label{df_f}
\end{equation} 
where $\delta\Omega = \Omega(A)-\Omega_0$ is the deviation of the amplitude-dependent resonance frequency $\Omega(A)$ from the unperturbed value $\Omega_0$, $m_{\text{eff}}$ is the effective mass of the mode and $A$ is the oscillation amplitude. So, the experimentally measured dependence of the resonance frequency on the amplitude, particularly its null result, may be used to set an upper limit for the model parameter $\beta_0$. 

The above mentioned theoretical considerations do not specify, which degree of freedom is subject to the QG corrections. If one considers a center of mass mode, then the scale of perturbation is strongly enhanced for the heavier than the Planck mass oscillators, as compared to individual atoms and molecules in the lattice. For instance, the precise measurement of the Lamb shift in hydrogen yielded an upper bound for the model parameter $\beta_0 < 10^{36}$~\cite{Das2008}. Although, the recent experiments with microscopic high-Q oscillators with effective masses ranging from $10^{-11}$~kg to $10^{-5}$~kg, established the new upper bound for $\beta_0 < 3 \times 10^{7}$~\cite{Marin2015}. The intrinsic acoustic nonlinearity of micro oscillators prevented to test quantum gravity corrections with the greater precision. 


In the following we describe an experiments with the sub-kilogram split-bar (SB) sapphire mechanical oscillator, where we demonstrate improvement for the upper value of the correction parameter $\beta_0$ compared to the previous work with intermediate range mechanical oscillators~\cite{Marin2015} by nearly an order of magnitude. In addition to the that, we provide the reasonable estimates of $\beta_0$ from experiments with bulk acoustic wave (BAW) quartz resonators yields the limit of $4\times10^4$. As the consequence of mean value entering in the right-hand side of the Eq.(1) the systems with higher mass and larger amplitude are preferred. As an example one may take measurements of the period of the physical pendulum in 1936~\cite{Atkinson1936}, where much lower upper bound of $\beta_0\lesssim 10^{-4}$ can be established from the deviation of the period dependence of amplitude from the well known Bernoulli non-linearity. However, due to the absence of the evaluation of the pendulum frequency stability the exact upper bound on $\beta_0$ can not be obtained.   

\section{III. MEASUREMENTS OF CORRECTION STRENGTH $\beta_0$ WITH SAPPHIRE DUMBBELL OSCILLATOR}

Microwave oscillators based on electromagnetic Whispering Gallery Mode (WGM) sapphire crystals offer excellent short- and middle-term frequency stability~\cite{Ivanov2009} due WGM high quality factors exceeding $10^8$ and existence of frequency-temperature turnover points. For these reasons these devices found applications in fundamental tests~\cite{Tobar2006,Giordano2012,Nagel2015}. The mechanical modes of sapphire resonators may attain $Q_M\simeq 10^8-10^9$~\cite{BraginskyBook,Tobar1997,Tobar2000}. The resonance frequencies of WGMs are very sensitive to changes in circumference, height of the cylinder resonator and to strain in the crystal lattice thus yielding the necessary coupling between mechanical and electromagnetic degrees of freedom for the observation of mechanical motion. Yet, no acoustic nonlinearities have been detected for the large sapphire mechanical resonators making these devices an excellent platform for QG tests.

\begin{figure}[ht!]
	\includegraphics[width=0.8\columnwidth]{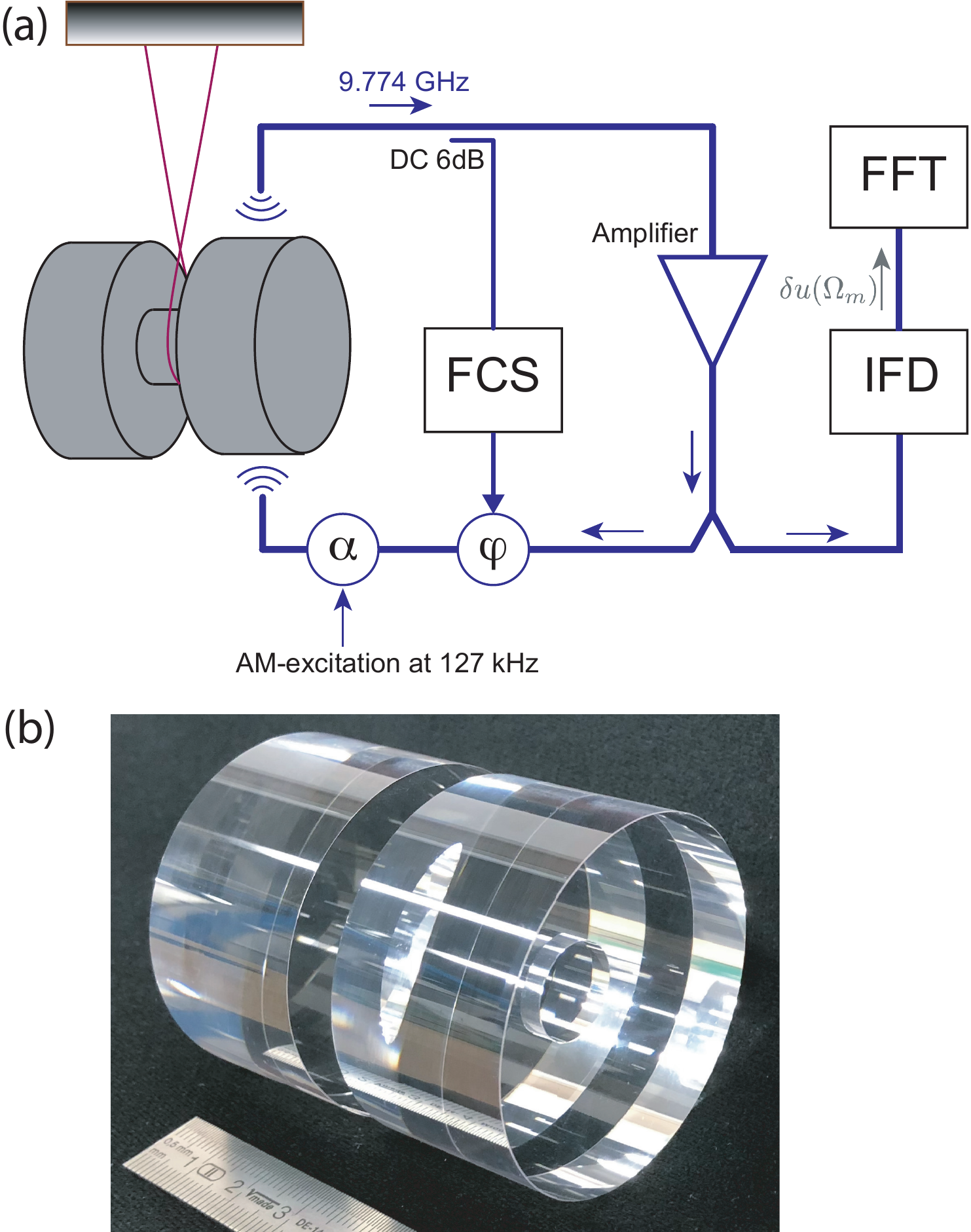}
	\caption{(Color online) (a) Simplified experiment schematic. See the text for details. (b) Picture of the sapphire SB resonator. The ruler shows the scale of the system. }\label{Sapphire_Oscillator}
\end{figure} 
   
The experimental setup, shown in Fig.~\ref{Sapphire_Oscillator}(a), is based on a cylindrical dumbbell shape or split-bar (SB) sapphire resonator, which is fabricated out of a single crystal HEMEX-grade sapphire fabricated by GT Advanced Technologies Inc., USA. The rotation symmetry axis of the resonator is parallel to the c-axis of the crystal. The SB resonator consists of two bars with diameter 55~mm and height 28~mm, which are separated by the neck of diameter 15~mm and length 8~mm, see Fig.~\ref{Sapphire_Oscillator}(b). Two electromagnetic WGM resonators are formed in each bars and undergo the same mechanical motion, i.e. they oscillate in phase for the breathing mode, which is similar to the fundamental longitudinal mode of the conventional cylindrical resonator of the same length and diameter. The resonance frequency of this mode is $\Omega_0/2\pi = 127.071~$kHz and its effective mass is calculated by using finite element modelling $m_{\text{eff}}=0.3$~kg. In order to maximize mechanical Q-factor, the resonator is suspended via a niobium wire-loop around the neck. The whole construction is placed inside temperature stabilized vacuum chamber at 300~K. The vacuum chamber in turn is placed on vibration isolation platform and kept at a pressure of $\sim10^{-2}$~mBar. 
 
A parametric transducer is used to detect the mechanical vibrations of the SB resonator, see ref.~\cite{Bourhill2015} for the details. For that purpose, the WGM sapphire resonator serves as a dispersive element inside a closed electronic loop, which together with an amplifier and a phase shifter constitute a microwave oscillator operating at the resonance frequency of the chosen WGM mode~\cite{Leeson1966}. An interferometric frequency control system (FCS) suppresses spurious phase fluctuations and locks the microwave oscillator to the frequency of the WGE$_{15,1,1}$ mode at $\omega_{\text{WGE}}/2\pi \simeq 9.774~$GHz~\cite{Ivanov1998}. The in-loop voltage-controlled attenuator $\alpha$ is used for the parametric excitation of the mechanical vibrations at 127~kHz. Approximately a half of the generated power inside the microwave sapphire oscillator is diverted to the interferometric frequency discriminator (IFD). The output signal of IFD is a linear function of its input frequency and is measured with HP 89410A spectrum analyzer. All instruments are time referenced to the hydrogen maser frequency standard VCH-103. 

The spatial overlap between microwave and mechanical modes results in the interaction between these degrees of freedom, which can be described by the standard opto-mechanical Hamiltonian $\hat{H}_{int}=-\hbar g_0 \hat{a}^{\dagger}\hat{a} \hat{x}$, where $g_0$ is a single photon opto-mechanical coupling, $\hat{a}^{\dagger},\hat{a}$ are raising and lowering operators for the WGM and $\hat{x}$ is canonical position operator for the center of mass mechanical motion~\cite{Aspelmeyer2014}. The microwave signal modulated at the resonance frequency of mechanical mode $\Omega_0$ induces radiation-pressure force which drives mechanical vibrations. The calibration of amplitude of center of mass motion is made by using the standard expression 
\begin{equation}
\delta u (\Omega)=\delta x (\Omega) (du/df) (df/dx), 
\label{dudx}
\end{equation}
where $df/dx$ is determined from the amplitude of the output IFD signal $\delta u(\Omega)$. That signal is proportional to the applied modulated power $\delta P$
\begin{equation}
\delta u (\Omega) = \chi \bigg( \frac{du}{df} \bigg) \bigg(\frac{df}{dx} \bigg)^2 \delta P,
\label{dudP}
\end{equation} 
where $\chi$ is the constant describing electromagnetic coupling of the signal and mechanical property of the oscillator~\cite{Bourhill2015}. The transduction constant is calculated to be $\delta x /\delta u = 526~$nm/mV. 
 
\begin{figure}[ht!]
	\includegraphics[width=0.8\columnwidth]{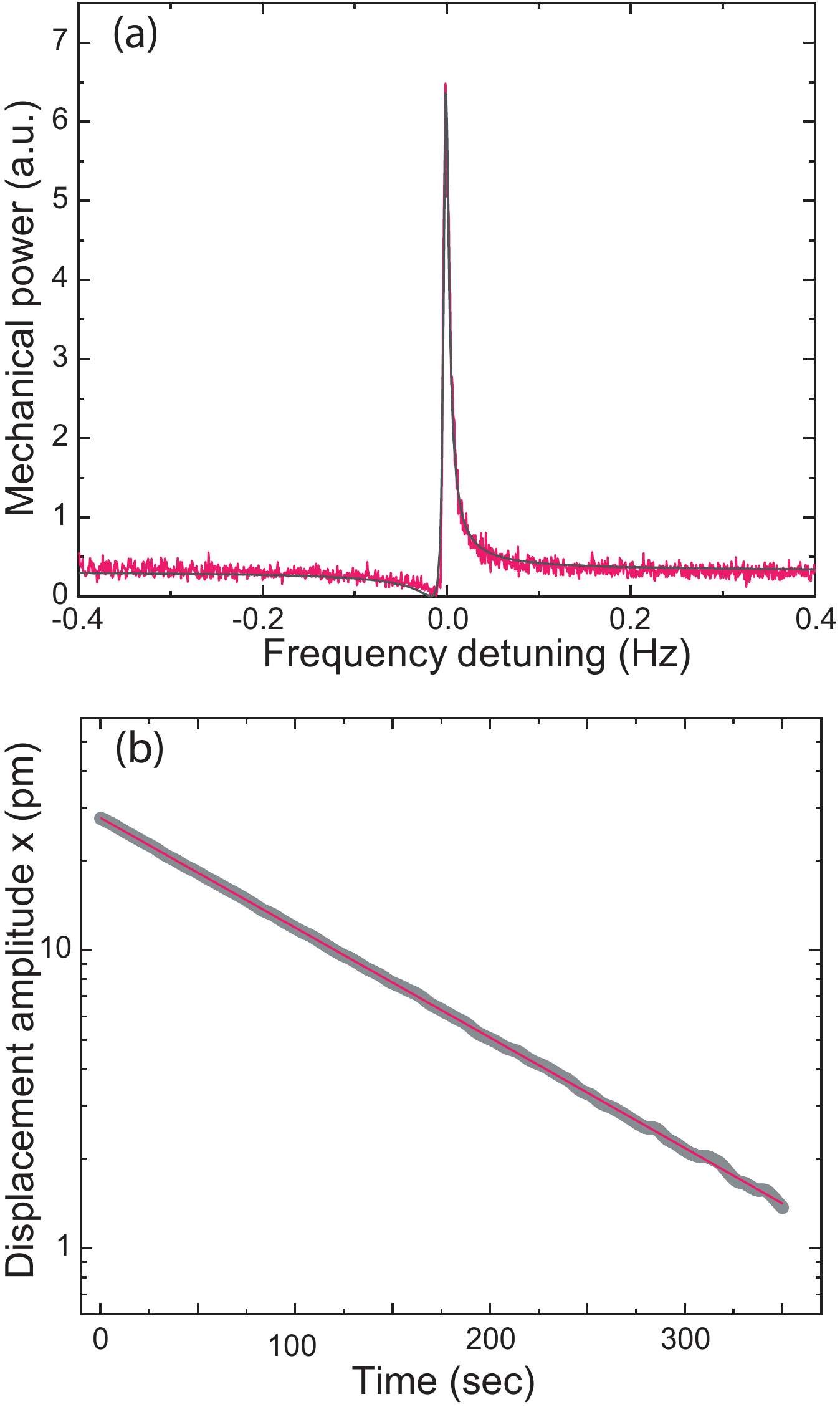}
	\caption{(Color online) (a) Mechanical response of the SB resonator to the acoustic excitation of the in-phase breathing mode around 127 kHz. The solid curve shows the fit to the quadrature signal due to double-balanced mixing. (b) Typical ringdown measurement of the in-phase breathing mode. The solid curve shows the fit to the exponential decay. }\label{RingDown}
\end{figure}

The mechanical response of the SB-resonator to the acoustic excitation in the vicinity of the resonance frequency is shown in Fig.~\ref{RingDown}(a). The applied excitation signal at 127~kHz is relatively weak resulting in the maximal amplitude of mechanical vibrations of 6~pm. The output signal is measured by using phase-sensitive interferometric setup which results in superposition of dispersive and absorptive quadrature components. The solid curve displays the fit of the experimental data to such composite absorptive-dispersive response and yields the resonance frequency of the mechanical resonator $\Omega_0/2\pi = 127070.9695 \pm 0.0003$~Hz, its FWHM linewidth $\Gamma_M/2\pi=3.5~$mHz and the 55$^{\circ}$ degrees mismatch between the arms of the IFD.  

The ringdown measurements of the mechanical vibrations are made in two steps. In the first step the resonance frequency of mechanical vibrations $\Omega_0/2\pi$ is determined. For that purpose, the radiation pressure force is applied to the resonator for the time sufficient to settle the mechanical vibrations (several minutes). Then, the output signal $\delta u(\Omega)$ is measured for every frequency point in the scanning range of 1 Hz. The resonance frequency corresponds to the point which yields the maximal IFD response $\delta u(\Omega)$. This procedure is repeated for the different excitation amplitudes (20-35 pm) in every single experimental run and detected no resonance frequency shift within
accuracy of 10 mHz determined by the resolution bandwidth of FFT analyzer. In the second step, after the mechanical resonance frequency is located, the AM-excitation is turned off, and then the mechanical vibrations are measured as they decrease due to acoustic losses. The amplitude and frequency of the decaying vibrations, i.e. the amplitude and the frequency of the spectral peak, is then tracked and recorded every 0.2~s. For that purpose a marker is placed on the maximum voltage value in the spectra, and its frequency and amplitude is recorded for every time bin. The frequency accuracy of such measurements is determined by the resolution bandwidth of the FFT analyzer, which is set to 5~Hz.

The typical ringdown measurement is presented in Fig.~\ref{RingDown}(b). For this particular example, the resonant frequency is $\Omega_0/2\pi=127070.97~$Hz. The solid curve shows the fit of the experimental data to the exponential decay with characteristic time constant $\tau=173~$sec which yields the mechanical quality factor $Q_M=\Omega_0\tau/2 = 3.4 \cdot 10^7$ and the same FWHM linewdith $\Gamma_M$ which is found in mechanical response measurements. In addition to the extracting of the parameters of the exponential decay, the Duffing equation was numerically solved in order to attain the best fit parameters for the ringdown amplitudes. Following this procedure we extract the upper limit for the QG model parameter to be $\beta_0<6\times10^{11}$. 

The frequency measurements yield much more stringent limit on $\beta_0$. In all measured ringdown series, there is no evidence of any detectable frequency shift up to the maximum amplitude of mechanical displacement of 75~pm. The null-frequency shift measured in the experiment corresponds to the accuracy of $ \delta \Omega /\Omega_0 = 3.9 \times 10^{-5}$ and accordingly to the Eq.~\ref{df_f} yields the upper limit for the QG model parameter $\beta_0<5.2 \times 10^6$. 

The sapphire SB resonator demonstrate a large potential for even more stringent test of $\beta_0 \ll 1$. Here, we propose two possible ways to improve the experiment. Firstly, the mechanical response (Fig.~\ref{RingDown}(a)) could be measured for much larger input power $\delta P$. It is possible to excite vibrations in sapphire resonator with amplitude of several nanometers~\cite{Tobar2000}. That would result is much higher signal-to-noise ratio and as a consequence would improve the accuracy of determination of the mechanical resonance frequency $\Omega_0$ to be better than $0.1~$mHz. Together with an increase of the oscillation amplitude up to $0.1-1$ nm scale, the upper limit on the correction strength may potentially be improved by at least 8 orders of magnitude arriving at the level $\beta_0 \lesssim 10^{-2}$. Secondly, one can implement an electromechanical sapphire oscillator by closing a feedback loop with the IFD output signal $\delta u(\Omega_M)$. In that case the uncertainty in determination of frequency shift will be decreased with the integration time $T$ as $1/\sqrt{\Omega_0 T}$. Assuming the driving amplitude of SB resonator of $A\simeq1~$nm and average time of $T=1~$ hour, the testable limit $\beta_0 \ll 1$ is within experimental accuracy.

\section{IV. ESTIMATION OF THE CORRECTION STRENGTH $\beta_0$ WITH BAW}

Another mechanical system, namely quartz bulk BAW resonator also constitutes a fruitful platform for precise tests of quantum gravity. This system exhibits high resonance frequency $\Omega_0/2\pi \simeq 10~$MHz, milligram scale of the effective mass of oscillating modes~\cite{Galliou2018}, large Q-factor close to $10^{10}$ at low temperatures~\cite{Goryachev2013} and high frequency stability of electromechanical oscillators reaching the level of $5 \times 10^{-14}$. The above listed features of quartz BAW are very attractive for fundamental tests such as Lorentz symmetry~\cite{Lo2016}. However, we note that quartz crystals possess its own quite strong elastic non-linearities that can mimic the quantum gravity effect. These non-linearities lead to a similar frequency shift, quadratic in amplitude and known as amplitude-frequency effect or isochronism, see ref.~\cite{GagnepainBook}, p. 245. This effect can be made to nearly vanish by means of an optimal choice of the cut angle of the crystal, known as LD-cut~\cite{Gufflet2001,Galliou2004}. The QG correction strength can be estimated from Eq.~\ref{df_f} and by using the experimental parameters $m_{\text{eff}}=5$~mg, $\Omega_0/2\pi=10~$MHz, $\delta \Omega/2\pi \simeq 1~$mHz, $A\simeq1~$nm. Our estimation yields $\beta_0 \lesssim 4 \times10^4$, which is still limited by elastic non-linearity. In order to single out quantum gravity frequency from such non-linearity, the amplitude frequency shift shall be measured in dependence on the effective mass of the resonating mode. We also believe that experimenting with kilogram scale quartz BAW~\cite{Vig2013} will result in much more stringent test of the quantum gravity model parameter in regime $\beta_0\lesssim 0.1$, because of weaker non-linearity due to the lower acoustic energy density and much larger effective mass. 

\section{V. ESTIMATION OF THE CORRECTION STRENGTH $\beta_0$ WITH  PHYSICAL PENDULUMS}

At present, the most stringent limit on correction strength $\beta_0$ is arguably set by the experiments with physical pendula, \textit{e.g.} by the 1936 Atkinson's measurements of the period of a kilogram-scale physical pendulum as a function of its amplitude. The combination of relatively large angular amplitudes, $\theta_0\sim1^{\circ}$ and large mass, $\sim1$~kg makes the pendulum an ideal object for testing the generalized commutator described by Eq.(\ref{eq:GUP}). Since the physical pendulum possesses an intrinsic softening non-linearity, leading to a quadratic (in the first order) dependence of the oscillation period on the angular amplitude, the QG correction can be considered as an additioal quadratic nonlinearity that can be determined from the accurate measurement of the period-amplitude dependence and comparison of the result against the well-known formula, see ref.~\cite{Landavshitz1}. Using Legendre polynomial approximation of the elliptic integral in the exact dependence of the pendulum period on angular amplitude $\theta_0$, one can rewrite Eq.(\ref{df_f}) for the physical pendulum as follows:
\begin{equation}
\frac{\delta T(\theta_0)}{T_0} = \Biggr[\frac{1}{16}-\beta_0 \bigg(\frac{2 \pi m L}{M_p c T_0}\bigg )^2\Biggl]\theta_0^2+\frac{11}{3072}\theta_0^4,
\label{df_f2}
\end{equation} 
where $\delta T(\theta_0)$ is amplitude dependent deviation of the period of the pendulum $T_0$, $m$ is the mass of the pendulum, $\theta_0 \simeq A/L$ is the angular amplitude of a pendulum and $L$ is its length. The numeric terms describe the intrinsic non-linearity of physical pendulum. The dependence between the rate and arc ($\theta_0$) for the free pendulum was measured already in 1936 ~\cite{Atkinson1936}, using a pendulum with length $L=1~$m and mass $m\simeq6~$kg. Using the data in this work, we estimate the $\beta_0\lesssim10^{-4}$. Other experiments with different kind of pendulums carried at different times shows no evidence of the deviation of the oscillation period from the conventional theory of physical pendulum~\cite{Smith1964,Fulcher1976}, and result in similar estimate on the upper limit for $\beta_0$~\cite{Plenio2019}. 

In general, mechanical pendulum cannot serve as an accurate clock, for its resonance frequency (or period) fluctuates over time due to various environmental factors: temperature, humidity, pressure, ageing, local gravity variation, seismic excitation \textit{etc}.  At best such system provides fractional frequency stability at $10^{-5}$ level at 1 sec, compared to $10^{-14}$ for the best quartz BAW. Spurious modes of the pendulum can also lead to significant cross talk between the modes, leading to unaccounted nonlinearities. The lack of data on such crucial metrological characteristics of the experiment as Allan deviation, i.e. the frequency stability of the oscillations, the Q-factor of a system, and the absence of any information on systematic errors ensuing from the design of the suspension point allows only a qualitative conclusion that $\beta_0\ll1$. For the proper measurement of $\beta_0$ one has to repeat the experiments with pendula or with metrological systems such as SB sapphire resonators or BAWs in a systematic metrological way.

\section{VI. CONCLUSION}

In this work, we designed an experiment to test the modification of dynamics of a mechanical oscillator following from the generalized uncertainty principle predicted by some phenomenological theories of quantum gravity. We set an upper limit on the QG correction strength $\beta_0$ to the canonical commutator \eqref{commutator1}, using an ultra-high-Q mechanical sapphire resonator with a sub-kilogram mass of the resonating mode. In the original work~\cite{Pikovski2012} an experiment is proposed where the sequence of light pulses separated by a quarter of a mechanical oscillation period is reflected off an oscillator four times before measuring its phase that should depend on $\beta_0$. Our results show that the same goal can be attained in a simpler and more reliable way using continuous RF measurements of frequency of the electro-mechanical oscillator that can be measured with higher precision compared to any other physical parameter.

\begin{figure}[ht!]
	\includegraphics[width=0.8\columnwidth]{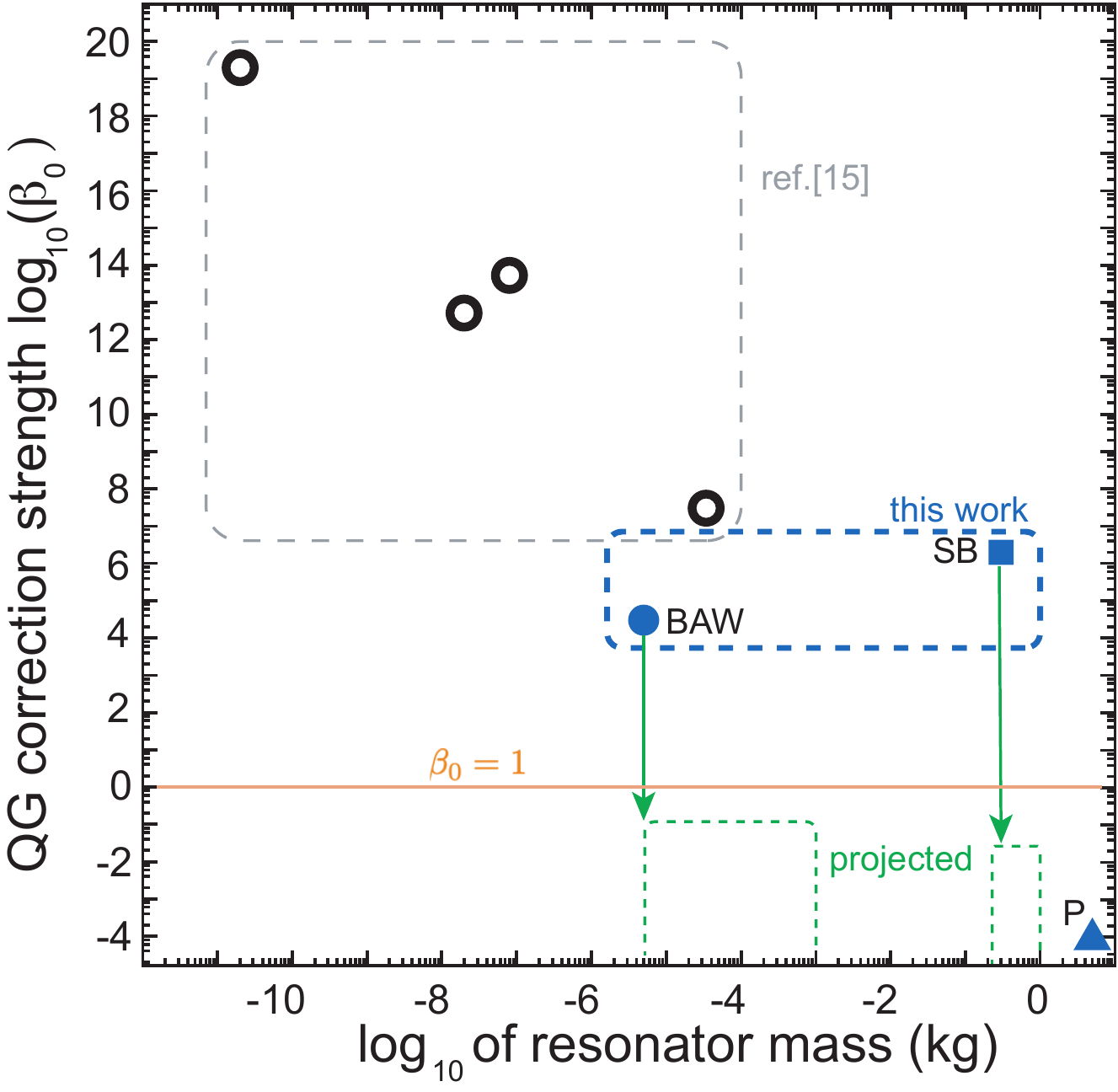}
	\caption{(Color online) The correction strength $\beta_0$ versus mass of mechanical oscillator determined in various experiments. The open circles correspond to $\beta_{0}$ reported in ref.~\cite{Marin2015}. The closed circle is the estimated $\beta_0$ for the quartz BAW at LD-cut, see ref.~\cite{Gufflet2001}. The square is the upper limit for $\beta_0$ obtained with sapphire split-bar resonator. The triangle shows the estimate of the correction strength from the measurements of the period of the physical pendulum, see ref.~\cite{Atkinson1936}.}\label{Beta}
\end{figure} 

The overview experimental tests of correction strength $\beta_0$ using mechanical oscillators is presented in Fig.~\ref{Beta}, where measured $\beta_{0}$ is plotted as a function of effective mass of the mechanical mode. Evidently, heavier oscillators result in more stringent limits on correction strength. Although massive mechanical pendula could be used to set a limit on $\beta_0\ll1$  in the variant of GUP relation described by Eq. \eqref{eq:GUP}, the high Q-factor and unprecedented frequency stability of the state-of-the-art quartz BAW resonators and SB sapphire resonators are more promising for precise tests of a more intuitive form of the GUP described by Eq.\eqref{GUP} that follows from the minimal length scale conjecture of quantum gravity theories ~\cite{Hossenfelder2013} as it requires a quantum level of sensitivity in frequency measurement. Having said that the perspective of utilizing of low-frequency ($<1~$Hz) mechanical oscillators or pendulums in this context remains unclear compared to the high-frequency system (kHz-GHz), where near quantum regime~\cite{LIGO2016} or even quantum limit~\cite{Kippenberg2009,Lehnert2009,Lehnert2011} has already been reached. 


\section{VII. ACKNOWLEDGMENTS}

The research was supported by the Australian Research Council Centre of Excellence for Engineered Quantum Systems CE170100009. PB thanks M. Plenio, R. Blatt, F. Scardigli, A. Vikman and P. Bosso for valuable discussions. SD would like to thank Lower Saxonian Ministry of Science and Culture that supported his research within the frame of the program “Research Line” (Forschungslinie) QUANOMET – Quantum- and Nano-Metrology. The authors are also very thankful to Y. Chen and the members of the MQM discussion group for insightful conversations that inspired this work.

\bibliography{QG}

\end{document}